\documentclass[runningheads]{llncs}
\usepackage{graphicx}
\usepackage{lscape}
\RequirePackage{rotating}
\usepackage{amsmath}
\usepackage{comment}
\usepackage{todonotes}
\usepackage{float}
\usepackage{censor}
\usepackage{rotating} 
\usepackage{siunitx} 
\usepackage[breaklinks=true,hidelinks]{hyperref}

\usepackage{xstring}

\usepackage[utf8]{inputenc}

\begin{document}
\title{Background Knowledge in Schema Matching: Strategy vs. Data}

\author{Jan Portisch\inst{1,2}\orcidID{0000-0001-5420-0663} \and
Michael Hladik\inst{3}\orcidID{0000-0002-2204-3138} \and
Heiko Paulheim\inst{1}\orcidID{0000-0003-4386-8195}}
\authorrunning{J. Portisch et al.}

\institute{Data and Web Science Group, University of Mannheim, Germany\\
\email{\{jan, heiko\}@informatik.uni-mannheim.de} 
\and
SAP SE Business Technology Platform | One Domain Model, Walldorf, Germany\\
\email{jan.portisch@sap.com}
\and
SAP SE Business Process Intelligence, Walldorf, Germany\\
\email{michael.hladik@sap.com}
}
\titlerunning{Background Knowledge in Schema Matching: Strategy vs. Data}

\maketitle
\begin{abstract}
The use of external background knowledge can be beneficial for the task of matching schemas or ontologies automatically. In this paper, we exploit six general-purpose knowledge graphs as sources of background knowledge for the matching task. The background sources are evaluated by applying three different exploitation strategies. 
We find that explicit strategies still outperform latent ones and that the choice of the strategy has a greater impact on the final alignment than the actual background dataset on which the strategy is applied. While we could not identify a universally superior resource, BabelNet achieved consistently good results.
Our best matcher configuration with BabelNet performs very competitively when compared to other matching systems even though no dataset-specific optimizations were made.

\end{abstract}

\section{Introduction}
\textit{Ontology matching} or \textit{schema matching} is the non-trivial task of finding correspondences between entities of two or more given ontologies or schemas. 
The matching can be performed manually or through the use of an automated matching system. In both cases, the context is very important and concept knowledge is required. Therefore, automated matching systems require background knowledge to excel at the schema matching task. In most cases, \textit{WordNet} is used as a form of general concept knowledge with a plain synonym lookup strategy. However, over the last decade many other sources of background knowledge that are much larger and also contain instance data have emerged. In addition, strategies to exploit structured knowledge, such as knowledge graph embedding models, have been developed but are rarely used in ontology matching. Exploiting background knowledge for ontology matching is still one of multiple challenges that is yet to be solved~\cite{challenges_2}. 

In this paper, we compare the performance of six different background data\-sets of varying size and characteristics for the task of schema matching. For each dataset, three different strategies are exploited. Besides an in-depth evaluation of the matching performance, we strive to test the 
following hypotheses:\\
\textbf{H1} The strategy is more important than the resource.\\
\textbf{H2} The resource is more important than the strategy.\\
\textbf{H3} There is a superior resource.\\
\textbf{H4} There is a superior strategy.

The remainder of this paper is structured as follows: In the next section, we present an overview on related work. Section~\ref{sec:approach} describes the general evaluation architecture that is used, as well as the generic matching process that was implemented for this paper. The background datasets and the strategies that are explored are presented in Sections~\ref{sec:data_sets} and \ref{sec:strategies}, respectively. The strategies on the background knowledge datasets are evaluated on four different gold standards in Section~\ref{sec:evaluation}. The paper closes with a summary and an outlook on future work.

\section{Related Work}
Ontology and schema matching systems are evaluated by the \textit{Ontology Alignment Evaluation Initiative (OAEI)}\footnote{\url{http://oaei.ontologymatching.org/}} every year since 2005.
While, to our knowledge, there is no large comparison of different general knowledge background sources or exploitation strategies, many individual matching systems exist that make use of external background knowledge.
In 2013, Euzenat and Shvaiko~\cite{euzenat_ontology_2013} counted more than 80 schema matching systems that exploit \textit{WordNet}. Besides WordNet, few other general background data sources are used: 
The \textit{WikiMatch}~\cite{shvaiko_wikimatch_2012} system exploits the \textit{Wikipedia} search API by determining concept similarity through the overlap of returned Wikipedia articles for a search term. \textit{WeSeE Match}~\cite{wesee_match_1} queries search APIs and determines similarity based on TF-IDF scores on the returned Web site titles and excerpts.
A synonymy and translation lookup strategy based on \textit{Wiktionary} is used in~\cite{DBLP:conf/semweb/PortischP20} for monolingual and multilingual matching. Lin and Krizhanovsky~\cite{lin_krizhanovsky} exploit \textit{Wiktionary} for translation lookups within a larger matching system.

In the biomedical and life science domain, specialized external background knowledge is broadly available and heavily exploited for ontology matching. Chen et al.~\cite{DBLP:conf/semweb/ChenXJC14} extend the LogMap matching system to use \textit{BioPortal}, a portal containing multiple ontologies, alignments, and synonyms, by (i) applying an overlap based approach as well as by (ii) selecting a suitable ontology automatically and using it as mediating ontology. As mappings between biomedical ontologies are available, those are used as well: Groß et al.~\cite{DBLP:conf/icbo/GrossHKR11} exploit existing mappings to third ontologies, so called \textit{intermediate ontologies}, to derive mappings. This approach is extended by Annane et al.~\cite{DBLP:conf/ekaw/AnnaneBAJ16} who use BioPortal by exploiting existing alignments between the ontologies found there for matching through a path-based approach: By linking source and target concepts into the global mapping graph, the paths that connect the concepts in that graph are used to derive new mappings. 
In the same domain, research has also been conducted on background knowledge selection. Faria et al.~\cite{faria2014automatic} propose the usage of a metric, called \textit{Mapping Gain (MG)}, which is based on the number of additional correspondences found given a baseline alignment. Quix et al.~\cite{DBLP:conf/sigmod/QuixRK11} use a keyword-based vector similarity approach to identify suitable background knowledge sources. Similarly, Hartung et al.~\cite{DBLP:conf/esws/HartungGKR12} introduce a metric, called \textit{effectiveness}, that is based on the mapping overlap between the ontologies to be matched. 
While in the biomedical domain, many specialized resources are available and data schemas are heavily interlinked, this is not the case for other domains. As a consequence, such methods cannot be easily translated and applied.

Background knowledge sources are also used for multilingual matching tasks. Here, translation APIs are often used such as \textit{Microsoft Bing Translator} by \textit{KEPLER}~\cite{kepler_oaei_18} or \textit{Google Translator} by \textit{LogMap}.

Approaches that exploit vector representations of concepts are rarely found in the ontology or schema matching domain. The \textit{DOME}~\cite{dome} matching system employs a \textit{doc2vec}~\cite{doc2vec} approach to concepts within the ontologies to be mapped. Similarly, \textit{AnyGraphMatcher}~\cite{agm} attempts to embed the ontologies to be mapped at runtime but achieves very low results in the OAEI 2019.
\emph{DESKMatcher}~\cite{DBLP:conf/semweb/MonychPHP20} applies a knowledge graph embedding approach on external knowledge but did not perform competitively in the OAEI 2020 either. \textit{WebIsAlod} is exploited as external background knowledge in~\cite{DBLP:conf/semweb/PortischHP20a} through a combined string matching and graph embedding strategy. 

These examples show that there is a larger body of works exploiting background knowledge with various strategies; however, they are always used in the context of a larger matching system. Ablation studies and therefore statements about the utility of a particular source and/or strategy are not available.

\section{General Approach}
\label{sec:approach}
To close this gap, we propose a simple, generic matching process that can work with different sources of background knowledge and exploitation strategies. Our aim is \emph{not} to build a top-performing matching system, but to provide a testbed for a fair comparison of different background knowledge sets and strategies.

\subsection{Overview}
\begin{figure}
    \centering
    \includegraphics[scale=0.25]{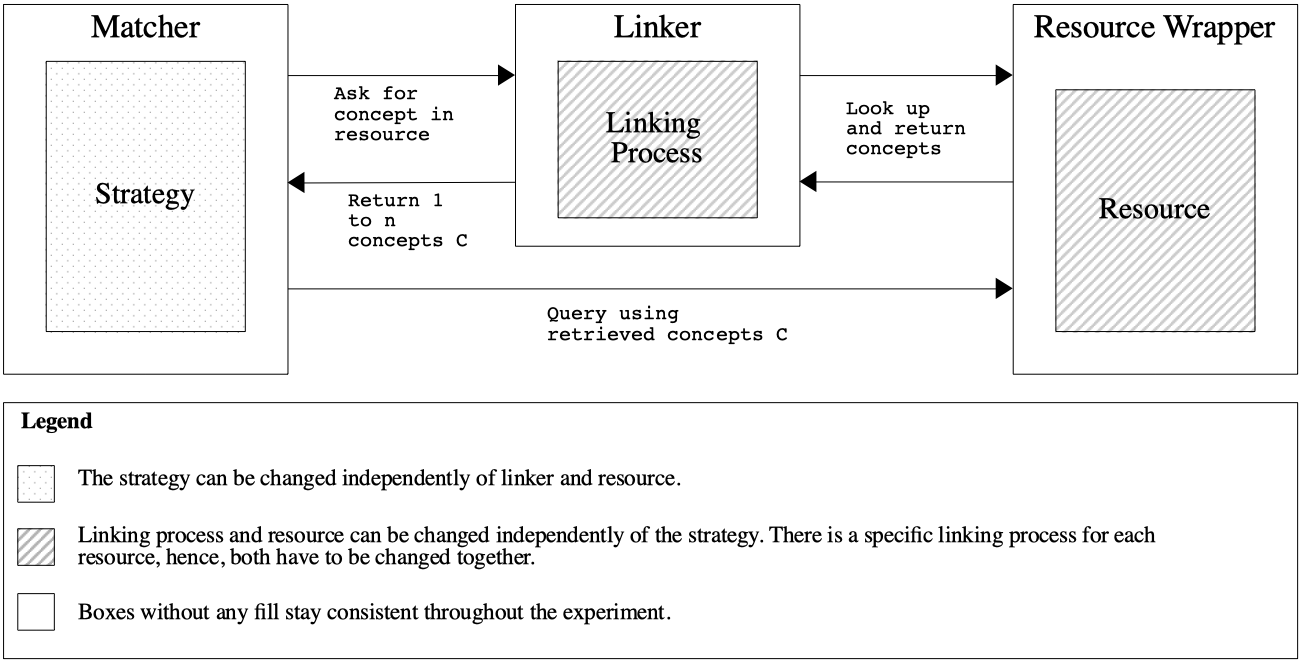}
    \caption{Architectural setting to evaluate different background datasets exploiting different strategies.}
    \label{fig:architecture}
\end{figure}

Figure~\ref{fig:architecture} depicts the architectural evaluation setting: A generic \textit{matcher} accepts two ontologies and outputs an alignment. Thereby, it applies a \textit{strategy} that can be exchanged independently of other matcher settings. Given labels, the matcher can ask a generic \textit{linker} whether a concept is available in a background knowledge source. Depending on the request type, the linker returns one or more corresponding concepts from the background knowledge. For \textit{Wiktionary}, for instance, the matcher can ask for concept \texttt{European Union} and the linker would return \texttt{dbnary-eng:European\_Union}. This linking process is also known as \emph{anchoring} or \emph{contextualization}~\cite{euzenat_ontology_2013_ch_7}.
Now that the matcher knows the representation in the background knowledge set, it can request further information through a generic \textit{resource wrapper} (such as similarities between concepts). Therefore, a \textit{resource} and a corresponding \textit{linking process} (that is wrapped by the \textit{linker}) have to be set. The implementation allows to change the \textit{resource} and the \textit{linking process} independently of other matcher settings such as the \textit{strategy}. 

\subsection{Matching Process}
\label{sec:matcher}
The matching process can be divided into two parts: linking and matching. The linking operation is implemented as a three step process: (i) \textit{Full Label Linking}, (ii) \textit{Longest Token Linking}, and (iii) \textit{Token Linking}. Later linking steps are only performed when the previous step was not able to link the label. In step (i), the full, i.e. unchanged, label is linked to a concept in the background knowledge source. 
Often, labels are composite concepts that do not appear in the knowledge source as a whole but in parts. To cover this case, step (ii) tokenizes labels and truncates them from the right. Linked parts are removed and the process is repeated to check for further concepts. This allows to detect long sub-concepts even if the full string cannot be found. Label \textit{conference banquet}, for example, cannot be linked to the Wiktionary background dataset using the full label. However, by applying right-to-left truncation, the label can be linked to two concepts, namely \textit{conference} and \textit{banquet}, and in the following also be matched to concept \textit{conference dinner} which is linked in the same fashion. The last fallback strategy is token linking (iii) which tokenizes each label (using spaces, underscores, and camel case recognition) and links the individual tokens to the background dataset.

After completion of the linking process, the match operation is performed. Multiple strategies are implemented here (see Section~\ref{sec:strategies}) which operate on the links. For the synonymy strategy, a match would be, for instance, annotated for \textit{(person, individual)} given that the two labels are synonymous according to the background dataset employed. If there are multiple links (linking steps (ii) and (iii)), a match requires that every link  has a matching partner (according to the strategy applied) in the set of links of the other label. In order to obtain a one-to-one alignment, the Hungarian extraction method~\cite{kuhn1955hungarian} is applied.

The overall matching runtime performance is improved by adding string matches directly to the final alignment. This step runs independently of the strategy or the background dataset used. It does not skew the outcome because all strategies under consideration in this paper are purely label-based. Hence, the same label used for two entities would always lead to a match.

Overall, the matching process scales with $O(nm)$ where $n$ is the number of elements in one ontology and $m$ is the number of elements in the other ontology.\footnote{The size of the external resource is not relevant within the matching process since all similarity functions applied here are lookup-based. When training an embedding with the external resource, the size of the resource affects scalability; however, the training is a one-time process -- once the vectors are available, they can be reused in all other matching tasks.}
It is important to note that the scalability can be improved by adding a candidate pre-selecion/blocking component. However, since scalability is not the main concern of this paper, we decided against complicating the matching pipeline.

The matcher is implemented using the \textit{Matching EvaLuation} \textit{Toolkit}~\cite{hertling_melt_2019,DBLP:conf/semweb/HertlingPP20} \textit{(MELT)}\footnote{\url{https://github.com/dwslab/melt/}}, an open-source Java framework for matcher development, tuning, evaluation, and packaging recommended by the OAEI. 
The matcher is implemented so that it is possible to use different sources of background knowledge and different strategies within the matching process. The implementation of this paper (linker, background sources, significance evaluation) has been unit tested, documented, and contributed to the framework so that other researchers can use the matching parts of the implementation (e.g. to easily use Wikidata synonyms/hypernyms through an API) for their matching system.\footnote{\url{https://dwslab.github.io/melt/matcher-development/with-background-knowledge}}

\section{Background Datasets}
\label{sec:data_sets}
For this paper, six knowledge graphs are exploited as background knowledge within the matching process. They are quickly introduced in the following:

\emph{BabelNet}~\cite{babelnet} is a large multilingual knowledge graph that integrates (originally) Wikipedia and WordNet. Later, additional resources such as Wiktionary were added. The integration between the resources is performed in an automated manner. The dataset does not just contain lemma-based knowledge but also instance data (named entities) such as the singer and songwriter \textit{Trent Reznor}. For the embedding strategy, the RDF version of BabelNet 3.6 was used\footnote{Unfortunately, there is no RDF version of the latest BabelNet version.}, for the other strategies, the BabelNet 4.1 indices. 

\emph{Wiktionary} is a ``collaborative project run by the Wikimedia Foundation to produce a free and complete dictionary in every language''\footnote{\url{https://web.archive.org/web/20190806080601/https://en.wiktionary.org/wiki/Wiktionary/}}. The project is organized similarly to Wikipedia: Everybody can contribute and edit the dictionary. The content is reviewed in a community process. Like Wikipedia, Wiktionary is available in many languages.

\emph{DBnary}~\cite{dbnary_2015} is an RDF version of Wiktionary that is publicly available.\footnote{\url{http://kaiko.getalp.org/about-dbnary/download/}} The DBnary dataset makes use of an extended \textit{LEMON} model~\cite{mccrae_interchanging_2012} to describe the data. For this work, a recent download from March 2021 of the English Wiktionary has been used. 

\emph{WebIsALOD} is a large hypernymy graph based on the \textit{WebIsA} database~\cite{webisa_db}. The latter is a dataset which consists of hypernymy relations extracted from the \textit{Common Crawl}, a large set of crawled Web pages. The extraction was performed in an automatic manner through Hearst-like~\cite{hearst} lexico-syntactic patterns. For example, from the sentence "[...] added that the country has favourable economic agreements with major economic powers, including the European Union.", the fact \texttt{isA(european\_union, major\_economic\_power)} is extracted.\footnote{\url{http://webisa.webdatacommons.org/417880315}} \textit{WebIsALOD}~\cite{webisalod} is the Linked Open Data endpoint which allows to query the data in SPARQL.\footnote{\url{http://webisa.webdatacommons.org/}} In addition to the endpoint, machine learning was used to assign confidence scores to the extracted triples. For this work, a confidence threshold of $0.5$ for hypernymy relations was chosen. The dataset of the endpoint is filtered, i.e. it contains a subset of the original \textit{WebIsA} database, to ensure a higher data quality. The knowledge graph contains instances as well as more abstract concepts that can also be found in a dictionary. 

\emph{WordNet}~\cite{fellbaum_wordnet} is a well-known and heavily used database of English words that are grouped in sets which represent one particular meaning, so called \textit{synsets}. The resource is strictly authored. \textit{WordNet} is publicly available, included in many natural language processing frameworks, and often used in research. An RDF version of the database is also available for download and was used for this work.\footnote{\url{http://wordnet-rdf.princeton.edu/about/}}

\emph{Wikidata}~\cite{vrandevcic2014wikidata} is a collaboratively built knowledge base containing more than 93 million data items. Like Wikipedia and Wiktionary, the project is run by the Wikimedia Foundation.
It is publicly available\footnote{\url{https://www.wikidata.org/wiki/Wikidata:Main_Page}} and under a permissive license. For this work, a download from March 2021 has been used.

\emph{DBpedia}~\cite{dbpedia} is a knowledge graph that is extracted from Wikipedia infoboxes. The underlying RDF files are available for download. For this work, the latest available files as of March 2021 have been downloaded via the DBpedia Databus\footnote{\url{https://databus.dbpedia.org/}} (rather than the 2016-10 version of DBpedia that is often used).

\section{Strategies}
\label{sec:strategies}
In the following, the exploitation strategies applied on the datasets outlined in the previous section are introduced.

\subsection{Synonymy}
\label{ssec:synonymy_strategy}
The synonymy strategy exploits existing synonymy relations in the datasets. On \textit{Wiktionary}, for instance, \textit{tired} is explicitly named as a synonym for \textit{sleepy}. 
Given two entities $e_1 \in O_1$ and $e_2 \in O_2$ of two ontologies $O_1$ and $O_2$, a match is annotated if the synonymy relation holds between at least one pair of their labels $l_{e_1}$ and $l_{e_2}$ according to the background dataset $B$ that is used. This is depicted in Equation~\ref{eq:synonymy}.
\begin{equation}
    isMatch_B(e_1, e_2) = isSynonymous_B(l_{e_1}, l_{e_2})
\label{eq:synonymy}
\end{equation}
The WebIsALOD dataset does not contain explicitly stated synonyms. Here, a synonym is assumed if both labels $l_{e_1}$ and $l_{e_2}$ appear as hypernyms of each other as shown in Equation~\ref{eq:synonymy_alod}. This occurs more often than one might assume due to the automatic extraction process that is applied to create this knowledge graph.\footnote{For example, \emph{symposium} and \emph{conference} are mutual hypernyms of each other in WebIsALOD.} The intuition behind the assumption here is that two things $X$ and $Y$ are describing the same thing if it was stated on the Web that $X$ is a $Y$ and that $Y$ is an $X$.
\begin{equation}
    isMatch_\text{WebIsALOD}(e_1, e_2) = isHypernymous(l_{e_1}, l_{e_2}) \land isHypernymous(l_{e_2}, l_{e_1})
\label{eq:synonymy_alod}
\end{equation}
For DBpedia, the properties \texttt{rdfs:label}, \texttt{foaf:name}, \texttt{dbo:alias}, \texttt{dbp:name}, and \texttt{dbp:otherNames} are used to obtain labels, and two entities are considered synonymous if they have at least one label in common. On Wikidata, we use \texttt{rdfs:label} and \texttt{skos:altLabel} to obtain labels, and determine synonymy with the same mechanism.

\subsection{Synonymy and Hypernymy}
\label{ssec:synonymy_hypernymy_strategy}
The synonymy and hypernymy strategy exploits the synonymy relations in the background datasets and, in addition, the hypernymy relations. Given two labels $l_{e_1}$ and $l_{e_2}$ of two entities $e_1$ and $e_2$, a match is annotated if one of the semantic relations holds between the two labels as depicted in Equation~\ref{equ:synonymy_and_hypernymy}.
\begin{equation}
\begin{split}
isMatch_B(e_1, e_2) = isSynonymous_B(l_{e_1}, l_{e_2})
\textrm{ }\\ \lor \textrm{ }isHypernym_B(l_{e_1}, l_{e_2})\textrm{ } \lor \textrm{ }isHypernym_B(l_{e_2}, l_{e_1})
\end{split}
\label{equ:synonymy_and_hypernymy}
\end{equation}
For DBpedia, the properties \texttt{rdf:type} and \texttt{dbo:type} are used to obtain hypernyms. On Wikidata, we use \texttt{wdt:P31} (instance of) and \texttt{wdt:P279} (subclass of).

\subsection{Knowledge Graph Embeddings}
\label{ssec:kge_strategy}
Knowledge graph embeddings, i.e. the vector-based representation of the elements within a knowledge graph, are a very active research area in recent years. Many such methods are known~\cite{DBLP:journals/corr/abs-1905-11485}. For this paper, we exploit the \textit{RDF2Vec}~\cite{rdf2vec_journal} approach: Random walks through the knowledge graph are generated starting from each node. The walks include the named edges of the graph. After the walk generation, the \textit{word2vec}~\cite{word_2_vec_2} algorithm is applied. Thereby, a vector representation for each node and each edge is obtained. This embedding approach has been chosen due to its simplicity, its good performance on a multitude of tasks (rather than being developed for only one task, RDF2Vec is task agnostic), its previous usage in ontology matching, and its scalability. It is important to note that the background knowledge source is transformed into a vector space -- not the ontologies that are to be matched.\\
Two entities $e_1 \in O_1$ and $e_2 \in O_2$ of two different ontologies $O_1$ and $O_2$ are matched if their labels $l_{e_1}$ and $l_{e_2}$ can be mapped to a vector $v_{l_{e_1}}$ and $v_{l_{e_2}}$ in the background knowledge dataset $B$ and the cosine similarity $sim$ between the two vectors is larger than a predefined threshold $t$. Hence: 
\begin{equation}
    isMatch_B(e_1, e_2) = sim(v_{l_{e_1}}, v_{l_{e_2}}) > t
\end{equation}
For WebIsALOD and WordNet, the pre-trained models from \textit{KGvec2go}\footnote{\url{http://kgvec2go.org/}}~\cite{kgvec2go} were used. The models were trained with the same configuration and, therefore, allow for comparability. Embeddings for the other three graphs are not available for download and were trained specifically for this paper.

Despite good scalability behavior of the embedding approach, vector representations for BabelNet, Wikidata, and DBpedia could not be calculated within 10 days. Therefore, \emph{RDF2Vec Light}~\cite{DBLP:conf/semweb/PortischHP20} was used for those very large knowledge graphs. The variant is based on the notion that, given a concrete task, only a small set of nodes within a knowledge graph are of actual interest. For example, given the matching task within the anatomy domain, a vector representation of \emph{Year Zero}, a music album by the industrial rock band \emph{Nine Inch Nails}, is not of particular interest. Therefore, a set of nodes of interest is defined in advance and walks are only generated for those. For ontology matching, the set of nodes of interest is known through the linking operation. Experiments showed that the performance of the light variant yields good results on various machine learning tasks compared to the classic variant~\cite{DBLP:conf/semweb/PortischHP20}.
For this work, the following parameters have been used: 500 walks per node, $depth=4$ (i.e., 4 node hops), \emph{SG} variant, $window=5$, and $dimension=200$. For the matcher configuration, a threshold of $t=0.7$ was used.

\subsection{Combination of Sources}
\label{ssec:combination_strategy}
The combination strategy exploits all datasets at the same time with the strategies mentioned above. For the synonymy strategy, a match is annotated if any background dataset finds evidence for a synonymy relation. The same logic is also applied in the synonymy and hypernymy strategy and the embedding strategy. 

\section{Evaluation}
\label{sec:evaluation}
We evaluate all combinations of the strategies presented in Section~\ref{sec:strategies} and background datasets presented in Section~\ref{sec:data_sets} on four evaluation datasets: (i) \textit{OAEI Anatomy} \cite{anatomy_original}, (ii) \textit{OAEI Conference} \cite{conference_original}, (iii) \textit{SAP FS}~\cite{banking_evaluation}, and (iv) \emph{LargeBio}. The experiments were performed on a 24 core server (à 2.6 GHz) with 386Gb of RAM running Debian 10.

\subsection{Evaluation Datasets}

Dataset (i) consists of two anatomical ontologies where the human anatomy has to be mapped to the anatomy of a mouse. The \textit{Conference} dataset (ii) consists of 16 ontologies from the conference domain and 120 alignment tasks between them. Out of those, 21 reference alignments are publicly available. The results reported in this paper refer to the available alignments. In order to allow for comparability with other matching systems, micro averages are reported; those are also reported by the \textit{OAEI Conference} track organizers. 
The \textit{SAP FS} dataset (iii) is a proprietary evaluation dataset from the banking and insurance industry consisting of 5 matching tasks. The ontologies in that dataset have been derived from conceptual data models. The dataset has been provided to the authors of this paper for research purposes by \textit{SAP SE Financial Services}. In order to allow for comparability with the numbers reported in the original paper, macro averages are reported here. From the LargeBio track (iv), the FMA/NCI small test case is used for the evaluation here. Overall, 21 matching system variants are evaluated on four tracks with a total of 28 test cases.

\subsection{Evaluation Metrics}
The alignments are evaluated using precision, recall, and $F_1$ which is the harmonic mean of the latter two. In addition, it is evaluated whether the alignments obtained by the different strategy-source combinations are significantly different. Therefore, a significance metric is required. For this work, we use McNemar's significance test as proposed by Majid et al.~\cite{DBLP:journals/tkdd/MohammadiAHT18}: Be $R$ the reference alignment and $A_1$, $A_2$ two system alignments. We can now calculate the two relevant elements from the contingency table as follows:

\begin{equation}
\begin{split}
    n_{01} = |(A_2 \cap R) - A_1| + |A_1 - A_2 - R|\\
    n_{10} = |(A_1 \cap R) - A_2| + |A_2 - A_1 -R|
\end{split}
\end{equation}

\noindent The significance can then be determined using McNemar's asymptotic test with continuity correction: 
\begin{equation}
\chi^2 = \frac{(|n_{01} - n_{10}|-1)^2}{n_{01} + n_{10}}
\end{equation}

\noindent For a small sample size ($n = n_{01}$ + $n_{10}$; $n < 25$), McNemar's exact test has to be used to obtain the p value: 

\begin{equation}
p = \sum^{n}_{x=n_{01}} (\binom{n}{x})(\frac{1}{2})^2
\end{equation}

\noindent For this paper, a significance level alpha of $\alpha = 0.05$ was chosen. As a side contribution of this work, the evaluation code for significance testing has been contributed to the MELT framework \cite{hertling_melt_2019} to facilitate reuse by other researchers.

\subsection{Results}
\begin{figure}[h]
    \centering
    \includegraphics[scale=0.24]{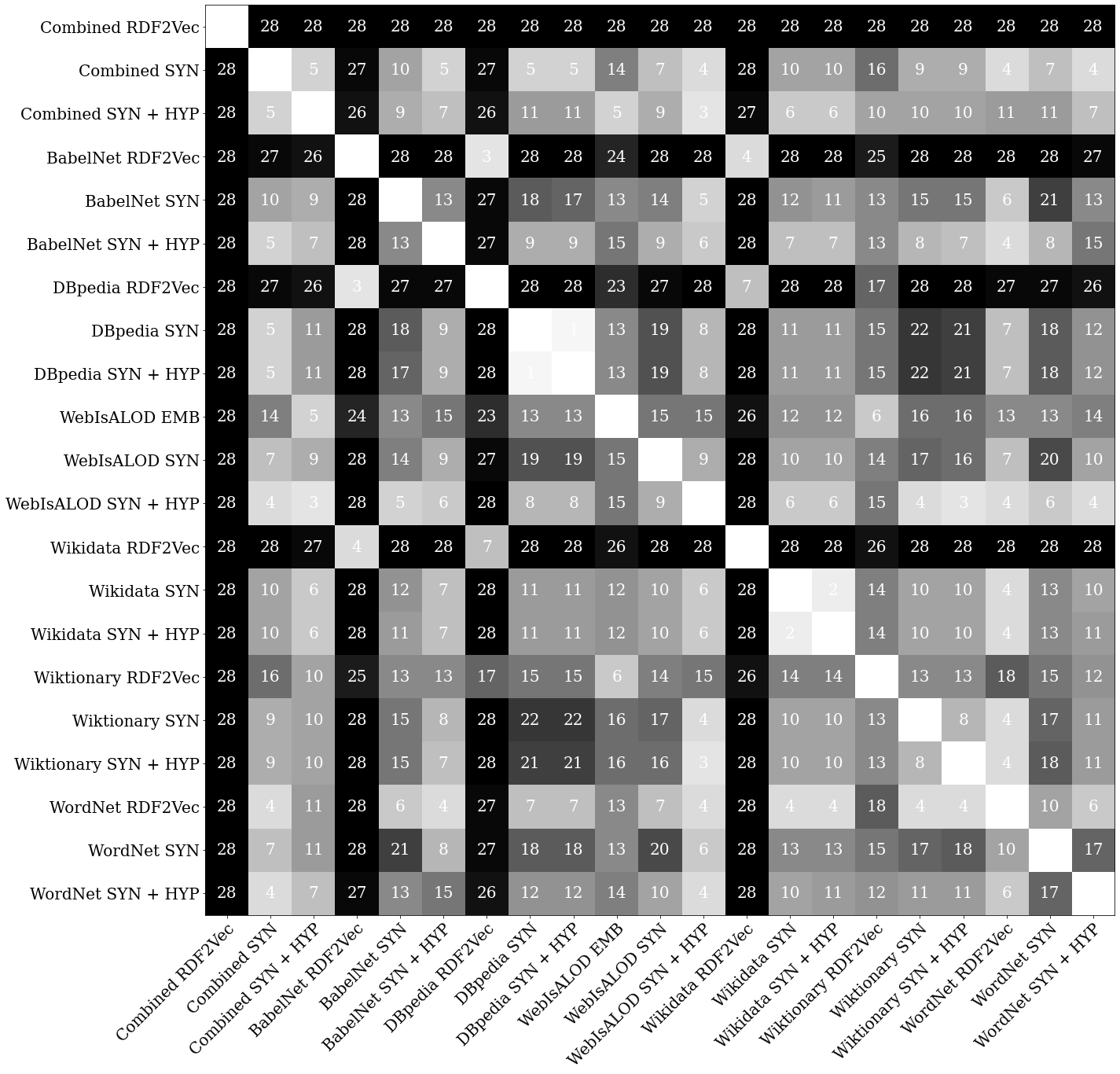}
    \caption{Matrix with the number of significantly different test case alignments given two matcher configurations. A higher total number of significantly different test case alignments has a darker shading in the figure. In total, there are 28 test cases.}
    \label{fig:all_syn_tc_significance}
\end{figure}

The performance results in terms of precision, recall, and $F_1$ are presented in Table~\ref{tab:evaluation}. The number of significantly different test case alignments is given in Figure~\ref{fig:all_syn_tc_significance}. More detailed performance and significance statistics as well as all alignments are available for download.\footnote{\url{https://github.com/janothan/bk-strategy-vs-data-supplements/}} It can be seen that the synonymy strategy consistently achieves the highest precision throughout all background knowledge resources. 
In terms of $F_1$, the synonymy strategy performs best in most cases when evaluating the strategy on each background source separately. The only area where the synonymy strategy falls short is recall. 
The significance tests show that despite similar scores, the alignments within this strategy group are significantly different in 285 out of 588 cases. This is also visible in Figure~\ref{fig:all_syn_tc_significance} which shows the number of significantly different alignments (given two matching systems). From the figure, it can be seen, for instance, that there are 22 significantly different alignments between DBpedia and Wiktionary using the synonymy strategy but only 5 different alignments between DBpedia and the combination approach using the synonymy strategy.

\begin{sidewaystable}[]
\caption{Evaluation results of six different background knowledge datasets exploiting three different strategies on four different gold standards. Note that for the \textit{Conference} task, micro averages are used while for the \textit{SAP FS} task, macro averages are reported. Baselines are given as reported by the OAEI organizers (FMA/NCI baselines are not provided).}
\begin{tabular}{ll|S|S|S|S|S|S|S|S|S|S|S|S|}
\cline{3-14}
 &  & \multicolumn{3}{c|}{\textbf{Anatomy}} & \multicolumn{3}{c|}{\textbf{Conference}} & \multicolumn{3}{c|}{\textbf{SAP FS}} & \multicolumn{3}{c|}{\textbf{\begin{tabular}[c]{@{}c@{}}FMA/NCI\\ (small)\end{tabular}}} \\ \hline
\multicolumn{1}{|c|}{\textbf{\begin{tabular}[c]{@{}c@{}}Knowledge\\ Graph\end{tabular}}} & \multicolumn{1}{c|}{\textbf{Strategy}} & \textbf{P} & \textbf{R} & \textbf{F1} & \textbf{P} & \textbf{R} & \textbf{F1} & \textbf{P} & \textbf{R} & \textbf{F1} & \textbf{P} & \textbf{R} & \textbf{F1} \\ \hline
\multicolumn{1}{|l|}{\textbf{BabelNet}} & \textbf{SYN} & 0.946843 & 0.75197889 & \multicolumn{1}{|r|}{\textbf{0.838}} & 0.67711 & 0.565895 & \multicolumn{1}{|r|}{\textbf{0.617}} & 0.40352626 & 0.15272657 & 0.2215869 & 0.909340659 & 0.3696947 & 0.52567496 \\ \hline 
\multicolumn{1}{|l|}{\textbf{}} & \textbf{SYN + HYP} & 0.9346405 & 0.754617 & 0.83503649 & 0.56206896 & 0.534426 & 0.547899159 & 0.358366758 & 0.15268351 & 0.214134 & 0.86925795 & 0.366344 & 0.515453 \\ \hline
\multicolumn{1}{|l|}{\textbf{}} & \textbf{\begin{tabular}[c]{@{}l@{}}RDF2Vec\\ (light)\end{tabular}} & 0.51411879259 & 0.6965699208 & 0.59159663 & 0.311879 & 0.26880119 & 0.28874225 & 0.2061385478 & 0.137710 & 0.16511569 & 0.311879 & 0.2688011 & 0.288742251 \\ \hline
\multicolumn{1}{|l|}{\textbf{WebIsALOD}} & \textbf{SYN} & 0.96682027 & 0.6919525 & 0.8066128 & 0.6593886 & 0.4950819 & 0.56554307 & 0.467575 & 0.143885038 & 0.22005382 & \multicolumn{1}{|r|}{\textbf{0.979}} & 0.26507818 & 0.417228 \\ \hline
\multicolumn{1}{|l|}{\textbf{}} & \textbf{SYN + HYP} & 0.91493055 & 0.695250659 & 0.790104947 & 0.457642 & 0.5372355 & 0.4942554 & 0.349247 & 0.14640464 & 0.2063198 & 0.86739659 & 0.265450483 & 0.4064994 \\ \hline
\multicolumn{1}{|l|}{\textbf{}} & \textbf{RDF2Vec} & 0.788575667 & 0.701187335 & 0.742318435 & 0.26872964 & 0.5409836 & 0.3590859 & 0.26073486 & 0.13992453405 & 0.1821158 & 0.63668430 & 0.2688011 & 0.37801047 \\ \hline
\multicolumn{1}{|l|}{\textbf{Wiktionary}} & \textbf{SYN} & \multicolumn{1}{|r|}{\textbf{0.968}} & 0.711741 & 0.820220 & 0.690677966 & 0.534426 & 0.6025878 & 0.4589141 & 0.145607 & 0.2210721 & 0.97677419 & 0.281831 & 0.43744582 \\ \hline
\multicolumn{1}{|l|}{\textbf{}} & \textbf{SYN + HYP} & 0.9668755 & 0.712401055 & 0.820357 & 0.674897119 & 0.5377049 & 0.598540145 & 0.4587493 & 0.145607834 & 0.2210530 & 0.97179487 & 0.282204 & 0.4373918061 \\ \hline
\multicolumn{1}{|l|}{\textbf{}} & \textbf{RDF2Vec} & 0.644016837 & 0.706464379 & 0.6737967 & 0.24883359 & 0.52459016 & 0.3375527 & 0.28574 & 0.14272146 & 0.19036204 & 0.47678795 & 0.282948622 & 0.3551401869 \\ \hline
\multicolumn{1}{|l|}{\textbf{WordNet}} & \textbf{SYN} & 0.9635555 & 0.7150395 & 0.8209011 & \multicolumn{1}{|r|}{\textbf{0.722}} & 0.527868 & 0.6098484 & 0.415209 & 0.148036455 & 0.2182567 & 0.96325 & 0.32204 & 0.48270089285714 \\ \hline
\multicolumn{1}{|l|}{\textbf{}} & \textbf{SYN + HYP} & 0.92875318 & 0.72222955 & 0.8126159 & 0.61363636 & 0.5311475 & 0.569420 & 0.3741607 & 0.151498959 & 0.2156717 & 0.87740628 & 0.3224125093 & 0.471549 \\ \hline
\multicolumn{1}{|l|}{\textbf{}} & \textbf{RDF2Vec} & 0.733287858 & 0.70910290 & 0.7209926 & 0.504672897 & 0.5311475 & 0.5311475 & 0.380115742875 & 0.1389549784 & 0.203513597 & 0.6257621951 & 0.30565897 & 0.41070535 \\ \hline
\multicolumn{1}{|l|}{\textbf{DBpedia}} & \textbf{SYN} & 0.935569285 & 0.699208 & 0.800302 & 0.70892 & 0.49508196 & 0.583011583 & 0.497262675 & 0.14222 & 0.2211920 & 0.9165067 & 0.35555 & 0.512339 \\ \hline
\multicolumn{1}{|l|}{\textbf{}} & \textbf{SYN + HYP} & 0.935569 & 0.699208 & 0.800302 & 0.70892 & 0.49508196 & 0.583011583 & \multicolumn{1}{|r|}{\textbf{0.537}} & 0.156059908 & \multicolumn{1}{|r|}{\textbf{0.242}} & 0.9165067 & 0.35555 & 0.512339 \\ \hline
\multicolumn{1}{|l|}{\textbf{}} & \textbf{\begin{tabular}[c]{@{}l@{}}RDF2Vec\\ (light)\end{tabular}} & 0.593905817 & 0.707124 & 0.645588678 & 0.1543756145526 & 0.514754098 & 0.2375189 & 0.224630890 & 0.1519484469 & 0.1812755 & 0.467644 & 0.344378257 & 0.39665523 \\ \hline
\multicolumn{1}{|l|}{\textbf{Wikidata}} & \textbf{SYN} & 0.924342105 & 0.7414248 & 0.8228404 & 0.636015 & 0.54426229 & 0.58657 & 0.44657 & 0.16317916 & 0.2390206 & 0.8941176 & 0.339538 & 0.4921748515 \\ \hline
\multicolumn{1}{|l|}{\textbf{}} & \textbf{SYN + HYP} & 0.9243421 & 0.74142 & 0.8228404 & 0.63601 & 0.544262295 & 0.58657 & 0.446360 & 0.163136 & 0.238943 & 0.89422135 & 0.3399106 & 0.4925816 \\ \hline
\multicolumn{1}{|l|}{\textbf{}} & \textbf{\begin{tabular}[c]{@{}l@{}}RDF2Vec\\ (light)\end{tabular}} & 0.5464720194 & 0.74076517 & 0.62895547 & 0.12453113278 & 0.54426229508 & 0.20268 & 0.2037966 & 0.1471509 & 0.17090225 & 0.354004 & 0.3209233 & 0.33665299 \\ \hline
\multicolumn{1}{|l|}{\textbf{Combinations}} & \textbf{SYN} & 0.881918819 & 0.788258 & 0.832462 & 0.49577464 & 0.57704918 & 0.53333 & 0.37601 & 0.1777831 & 0.24141958 & 0.8290155 & 0.476545 & \multicolumn{1}{|r|}{\textbf{0.605}} \\ \hline
\multicolumn{1}{|l|}{\textbf{}} & \textbf{SYN + HYP} & 0.81509177 & \multicolumn{1}{|r|}{\textbf{0.791}} & 0.8028 & 0.338951 & 0.5934426 & 0.4314660 & 0.30123 & 0.17875 & 0.22436515 & 0.727477 & \multicolumn{1}{|r|}{\textbf{0.481}} & 0.5791125 \\ \hline
\multicolumn{1}{|l|}{\textbf{}} & \textbf{RDF2Vec} & 0.24105 & 0.78627 & 0.36898 & 0.04960 & \multicolumn{1}{|r|}{\textbf{0.620}} & 0.24183006 & 0.13893309 & \multicolumn{1}{|r|}{\textbf{0.194}} & 0.162076 & 0.160778 & 0.4184661 & 0.2323033 \\ \hline
\multicolumn{1}{|l|}{\textbf{Baseline}} & \multicolumn{1}{c|}{-} & 0.997 & 0.622 & 0.766 & 0.76 & 0.41 & 0.53 & 0.52 & 0.15 & 0.23 & \multicolumn{1}{c|}{-} & \multicolumn{1}{c|}{-} & \multicolumn{1}{c|}{-} \\ \hline
\end{tabular}
\label{tab:evaluation}
\end{sidewaystable}

With the exception of BabelNet, the addition of hypernyms increases recall.\footnote{This may seem odd at first. However, lower recall values are due to the Hungarian optimization method to obtain a 1:1 alignment, which, in that case, extracts more false positives.} However, a drop in precision leads to overall lower $F_1$ scores (with the exception of DBpedia on SAP FS and Wikidata on FMA/NCI). The results indicate that hypernyms could be used in more complex matching strategies, e.g. as part of candidate generation. Nonetheless, a na\"{i}ve merge of synonymy and hypernymy sets as main strategy is not generally suitable for precise matching on the given evaluation datasets.

The embedding-based approach falls short of performing competitively. While the recall can be increased in some cases, the method generally scores a significantly lower precision leading to an overall low $F_1$ score.
One likely reason for the bad performance of the embeddings is that the RDF2Vec vector similarity seems to be an indicator for relatedness rather than actual concept similarity -- an observation that has also been made earlier~\cite{kgvec2go}. More promising usage scenarios for the embedding models exploited in this paper are likely candidate selection and hybrid strategies. Concerning significance testing, the embedding strategies produce the most significantly different alignments of all strategies evaluated in this paper. 
In addition, it was observed that embedding large background knowledge datasets is computationally very expensive which does not apply to the matching run time after the models were trained. 

Concerning the choice of background knowledge, WordNet, Wiktionary, and BabelNet are similar in the sense that they are focused on lexical facts. BabelNet, the largest of the three, scores the overall best $F_1$ score on Anatomy and Conference. On the remaining two tracks, the performance is competitive.

Despite its small size, WordNet also achieves competitive results compared to Wiktionary on Anatomy, Conference, and SAP FS and outperforms the latter significantly on the LargeBio task. Nevertheless, unlike WordNet, Wiktionary and BabelNet are constantly growing over time due to a community-driven creation process and might outperform WordNet in the long run.

DBpedia performs in the mid-range in terms of $F_1$. The recall is lower than that of the better performing systems (BabelNet, Wiktionary, WordNet). The most likely explanation is a lower concept coverage since DBpedia contains rather instances than class concepts. Interestingly, the addition of hypernyms has rarely any effect on this particular background source.

Wikidata performed similarly to DBpedia. Like the latter dataset, the addition of hypernyms does not change the results significantly. 

The WebIsALOD dataset achieves the lowest overall results. The most likely reason is that the dataset is not authored but automatically built leading to a lot of noise contained in the dataset (wrong hypernyms). The comparatively bad performance of the synonymy strategy may be grounded in the fact that WebIsALOD is the only graph evaluated here that does not explicitly state synonyms -- but instead those are derived, as outlined before, which is less precise.

The combination of different background knowledge sources increases the recall in all cases. Except on the LargeBio dataset, the drop in precision cannot make up for increases in recall.

When comparing the performance numbers on evaluation dataset level, it can be seen that the \textit{Anatomy} matching task achieves the best results -- this is likely due to a high textual overlap of the labels. On the \textit{Conference} task, the matchers achieve a lower precision and recall score. These observations are in line with those at \textit{OAEI} campaigns. On the domain specific \textit{SAP FS} dataset, it can be seen that recall and precision scores are low. Likely explanations here are a domain specific vocabulary, low explicitness of knowledge (many semantic details are hidden in lengthy descriptions) as well as a complex many-to-many matching problem (see \cite{banking_evaluation} for details).   

\begin{figure}
    \centering
    \includegraphics[scale=0.32]{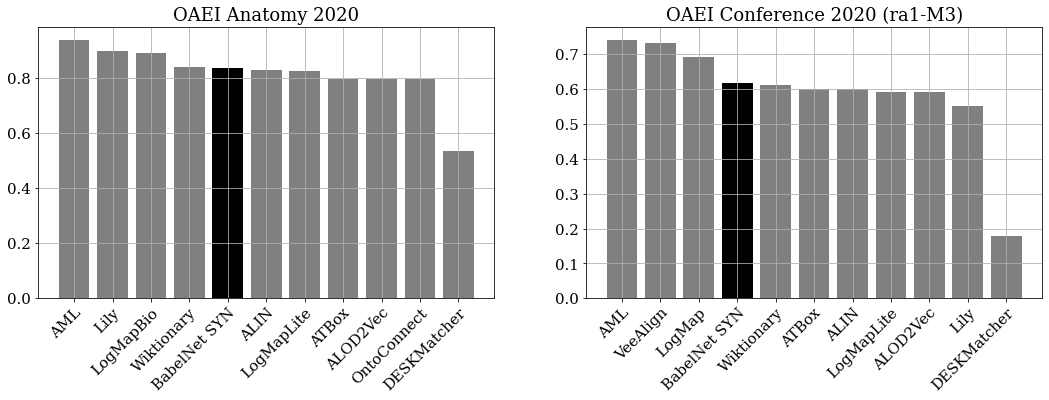}
    \caption{Performance in terms of $F_1$ on the OAEI Anatomy and Conference tracks of 2020.}
    \label{fig:oaei_performance}
\end{figure}

It is important to note that the work presented here is not intended to be a full-scale matching system but rather a comparison of different background knowledge datasets and exploitation strategies. Nonetheless, the performance of the best matching results achieved here on \textit{Anatomy} and \textit{Conference} are comparable to \textit{OAEI} matching results reported in the most recent 2020 campaign. A comparison in terms of $F_1$ is depicted in Figure~\ref{fig:oaei_performance}. It can be seen, that the best configuration of this paper performs in both cases above the median of the systems submitted in 2020. On Anatomy, it is noteworthy, that the first three systems (AML, Lily, and LogMapBio) use domain-specific resources leading to an advantage over the general-purpose resources exploited in this work.

\subsubsection{Hypotheses}
In order to evaluate hypotheses 1 and 2, we averaged the relative share of significantly different alignments on all test cases (i) while keeping the background source constant and changing the strategy (Equation~\ref{eq:impact_strategy}) and (ii) while keeping the strategy constant and changing the background source (Equation~\ref{eq:impact_bk}):

\begin{equation}
    impact_{strategy} = \frac{\sum_{bk \in BK}  \frac{ \sum_{tc \in TC} \sum_{s_1 \in S} \sum_{s_2 \in S} sig(m(bk, s_1), m(bk, s_2))  }{|TC| * |S|^2 - |TC| * |S|}}{|BK|}
    \label{eq:impact_strategy}
\end{equation}

\begin{equation}
    impact_{source} = \frac{\sum_{s \in S}  \frac{ \sum_{tc \in TC} \sum_{bk_1 \in BK} \sum_{bk_2 \in BK} sig(m(bk_1, s), m(bk_2, s))  }{|TC| * |BK|^2 - |BK| * |S|}}{|S|}
    \label{eq:impact_bk}
\end{equation}

\noindent where $S$ is the set of strategies, $BK$ is the set of background sources, $sig(align\-ment_1, alignment_2)$ is the significance function which will return 1 if the two provided alignments are significantly different and else 0, and $m(bk, s)$ is the matching function which returns the alignment by using the specified background knowledge source $bk$ and strategy $s$.

While keeping the background knowledge source constant and changing the strategy, we observed on average 57.5\% significantly different alignments with a standard deviation of $\sigma = 0.163$. On the other hand, while keeping the strategy constant and changing the background knowledge source, we obtained on average 51.76\% significantly different alignments with a standard deviation of $\sigma = 0.181$. Given our experimental setup, we hence accept H1 and reject H2 since a variation in the strategic component has a higher impact on the alignments than a variation of the background sources under consideration in this study. It is noteworthy that both components lead on average to more than 50\% significantly different alignments.
Since our results do not indicate that there is a superior resource over all test sets, we can reject H3. However, it is noteworthy that BabelNet achieves consistently good (on two tracks the best) results in terms of $F_1$ when using the synonymy strategy. 
Similarly, we do not find a superior strategy over each and every single test case and reject H4 -- but yet, the synonymy strategy achieved the best $F_1$ score on 3 out of 4 tracks and consistently performed very well compared to the other strategies.

\section{Summary and Future Work}
\label{sec:summary_future_work}
In this paper, we evaluated three different matching strategies using six different general purpose knowledge graphs on various evaluation datasets. 
We find that the strategy influences the final alignment more than the underlying dataset. 
Given the strategies evaluated here, those exploiting explicitly stated knowledge outperform a latent strategy. However, the exploitation of graph embeddings for data integration and schema matching is novel and its performance is still very low. 
While no superior general knowledge dataset could be identified, BabelNet produced consistently good or the best results. The humanly verified datasets outperformed the automatic generated one. Concerning the level of authoring between the datasets, the results indicate no clear superiority of expert-validated knowledge graphs over those created and validated by an open community.

In the future, we plan to exploit further embedding strategies, such as translational approaches, for schema matching as well as graph-based and dataset specific strategies. We further plan to examine more domain-specific matching tasks such as the \textit{SAP FS} dataset.

\bibliographystyle{splncs04}
\bibliography{references}

\end{document}